\documentclass[twocolumn,aps,superscriptaddress]{revtex4}
\usepackage{amssymb}
\usepackage{amsmath}
\usepackage{graphicx}
\usepackage[normalem]{ulem}
\usepackage[dvips]{color}
\usepackage{appendix}

\usepackage{color}
\usepackage[normalem]{ulem}

\begin{document}

\title{Extracting temperature of plasma through fusion reactions within a transport approach}
\author{Zhe Zhu}
\affiliation{Shanghai Institute of Applied Physics, Chinese Academy
of Sciences, Shanghai 201800, China}
\affiliation{University of Chinese Academy of Sciences, Beijing 100049, China}
\affiliation{Cyclotron Institute, Texas A$\&$M University, College Station, TX, USA}
\author{Jun Xu}\email[Correspond to\ ]{junxu@tongji.edu.cn}
\affiliation{School of Physics Science and Engineering, Tongji University, Shanghai 200092, China}
%\affiliation{Shanghai Advanced Research Institute, Chinese Academy of Sciences, Shanghai 201210, China}
\affiliation{Shanghai Institute of Applied Physics, Chinese Academy of Sciences, Shanghai 201800, China}
\date{\today}

\begin{abstract}
We have investigated the method of extracting the temperature from weighted proton-to-neutron yield ratio from fusion reactions as in the previous experiment~[W. Bang, {\it et al.}, Phys. Rev. Lett. \textbf{111}, 055002 (2013).] using the Texas Petawatt laser beam. The Coulomb explosion of deuterium clusters is simulated based on the particle-in-cell model in a box system with periodic boundary conditions, and fusion reactions are incorporated through the stochastic method. As long as the deuteron numbers in deuterium clusters follow a log-normal distribution, the low-energy part of the final deuteron spectrum can be fitted by a Maxwell-Boltzmann distribution, while there are more deuterons in the intermediate- and high-energy region compared to a thermal distribution, and dominate the weighted yield ratio. Therefore, the effective temperature extracted from the weighted yield ratio is generally higher than that from fitting the final deuteron spectrum. The local density fluctuation, which intrinsically exists due to the log-normal distribution of deuteron numbers, further enhances hot deuteron-deuteron collisions, and significantly affects the weighted yield ratio.
\end{abstract}

\maketitle

\section{Introduction}
\label{sec:intro}

Nuclear fusion reactions in the plasma environment induced by high-intensity laser beam have been a hot research field in the past two decades~\cite{Dit98,Smi00}, thanks to the fast development of chirped pulse amplification technology. Initially, such reactions, e.g., D(d,n)$^3$He on targets formed of deuterium clusters, were considered to be a potential neutron source~\cite{Dit99}, which may have wide applications in other research fields, e.g., material science. Later on, attempts were made to increase the neutron yield by using different targets~\cite{Gri02,Mad04}. The scaling relation between the neutron yield and the laser pulse energy was further analyzed~\cite{Won23}. Since the temperature reached in the plasma is similar to that in early Universe or inside stars, such processes provide useful ways of studying the primordial nucleosynthesis on earth. For example, with the help of the Texas Petawatt laser, the temperature of the system was measured from the density weighted yield ratio of protons to neutrons produced, respectively, in $^3$He(d,p)$^4$He and D(d,n)$^3$He reactions, which is somehow slightly higher than the temperature extracted from the time-of-flight distribution of deuterium ions representing their kinetic energy spectrum~\cite{Ban13}. With the extracted temperature, the $S$ factor of the $^3$He(d,p)$^4$He reaction at low center-of-mass (C.M.) energies was further obtained~\cite{Bar13}.

Understanding the dynamics and properties of plasma system where nuclear fusion reactions take place is important in extracting reliable information from experiments mentioned above. The Coulomb explosion model~\cite{Kra02} is widely used to describe the dynamics of the system. While the energy distribution of the ions driven by the electrostatic field may resemble a Maxwell-Boltzmann (MB) form as a consequence of the size distribution of clusters~\cite{Kra02,Dit99,Zwe02}, the ions are not necessarily thermalized and, as such, the extraction of the temperature based on the assumption of the MB distribution as in Refs.~\cite{Ban13,Bar13} is an approximation. Indeed, one can estimate that the lifetime of the system is much shorter compared to the relaxation time for thermal equilibrium under short-range Coulomb collisions~\cite{Kra02,Spi67,Smi01}. In order to describe the dynamics in such non-equilibrium systems, various theories~\cite{Bre05} and models~\cite{Las04,Hei09,Che15,Arb15} have been developed. Since the molecular dynamics model~\cite{Las04,Hei09,Che15} is only suitable for the Coulomb explosion of small clusters within limited computational power, we employ the particle-in-cell model EPOCH~\cite{Arb15}. In our previous study~\cite{Zhu22}, we have incorporated the D(d,n)$^3$He channel into EPOCH through the stochastic method~\cite{Xu05}, and have studied the relation between the neutron yield and the energy distribution of deuterons from the Coulomb explosion of deuterium clusters in a system with free boundary conditions. In the present study, we incorporate more reaction channels with the similar stochastic method, and simulate the dynamics as in Ref.~\cite{Ban13} in a globally charge-neutral box system with periodic boundary conditions. We find that the widely-used log-normal distribution of deuteron numbers in deuterium clusters does lead to a MB-like kinetic energy distribution at low energies, while fusion reactions are dominated by deuterons at intermediate and high energies and are significantly affected by local density fluctuations.

\section{Theoretical framework}
\label{sec:theory}

While EPOCH is a typical particle-in-cell transport model, with each simulation particle representing a number of real particles, we set the weight of deuterons to be 1 in order to properly simulate the Coulomb explosion of deuterium clusters. These charged particles propagate under the electromagnetic (EM) field and generate currents~\cite{Esi01,Vil92}, and the EM field is generated by solving Maxwell's equations based on the current on a fixed spatial grid~\cite{Yee66,Bor70}. As long as the particles in EPOCH consistently evolve under the EM field via current deposition, the Gauss law, i.e., $\nabla \cdot \vec{E}= \rho_q/\epsilon_0$ with $\vec{E}$, $\rho_q$, and $\epsilon_0$ being, respectively, the electric field, the charge density, and the dielectric constant, is always satisfied with proper initialization. A constant time step of $1.6\times10^{-3}$ fs is used for the calculation of the EM field. In order to increase the numerical accuracy in solving the differential equation and avoid self-heating, we employ a 5th-order B-spline method for the shape function of each simulation particle. Besides the mean-field evolution under the EM field, charged particles may also experience Coulomb collisions by using the approach in Refs.~\cite{Sen98,Sen08}, where a particle can only collide with another particle in the same cell, and the collision algorithm is executed in each cell in the simulation area. This approach treats short-range Coulomb collisions stochastically in momentum space, with the energy conserved perfectly in each collision but momentum conserved on average. For a recent improved treatment on Coulomb collisions, see Ref.~\cite{Hig20}. For more details of EPOCH, we refer the reader to Ref.~\cite{Arb15}.

As estimated in Ref.~\cite{Que18}, the pulse duration of the Texas Petawatt laser is longer than the disassembly time of a single cluster but shorter than the time for the mixture of deuterons from different clusters through their Coulomb explosions. In the present study we neglect the laser interaction during the Coulomb explosion and assume that all deuterium clusters are ionized. In contrast to our previous work~\cite{Zhu22}, we employ periodic boundary conditions in the present study, with initially deuterium clusters of different sizes and uniformly distributed electrons forming a globally charge-neutral plasma system. In the cubic box system, a particle that escapes from one side will enter the box from the opposite side with the same velocity. The shape functions of charged particles across the boundary of the box also expand to the opposite side of the box and are properly treated. Since the evolution of the EM field is iterated from the initial state, the initial electric field is properly prepared to suit periodic boundary conditions. To ensure the energy conservation, the electric potential should be continuous at boundaries of the box in a periodic manner, and this is achieved by constructing images of the electric field generated by the plasma system periodically in space. The grid length for the EM field calculation is $\Delta r = 8.81\times10^{-4}$ $\mu$m, and the size of the box ranges from $(351 \Delta r)^3$ to $(450 \Delta r)^3$ for 100 simulation events in total.

\begin{figure}[t]
\includegraphics[scale=0.3]{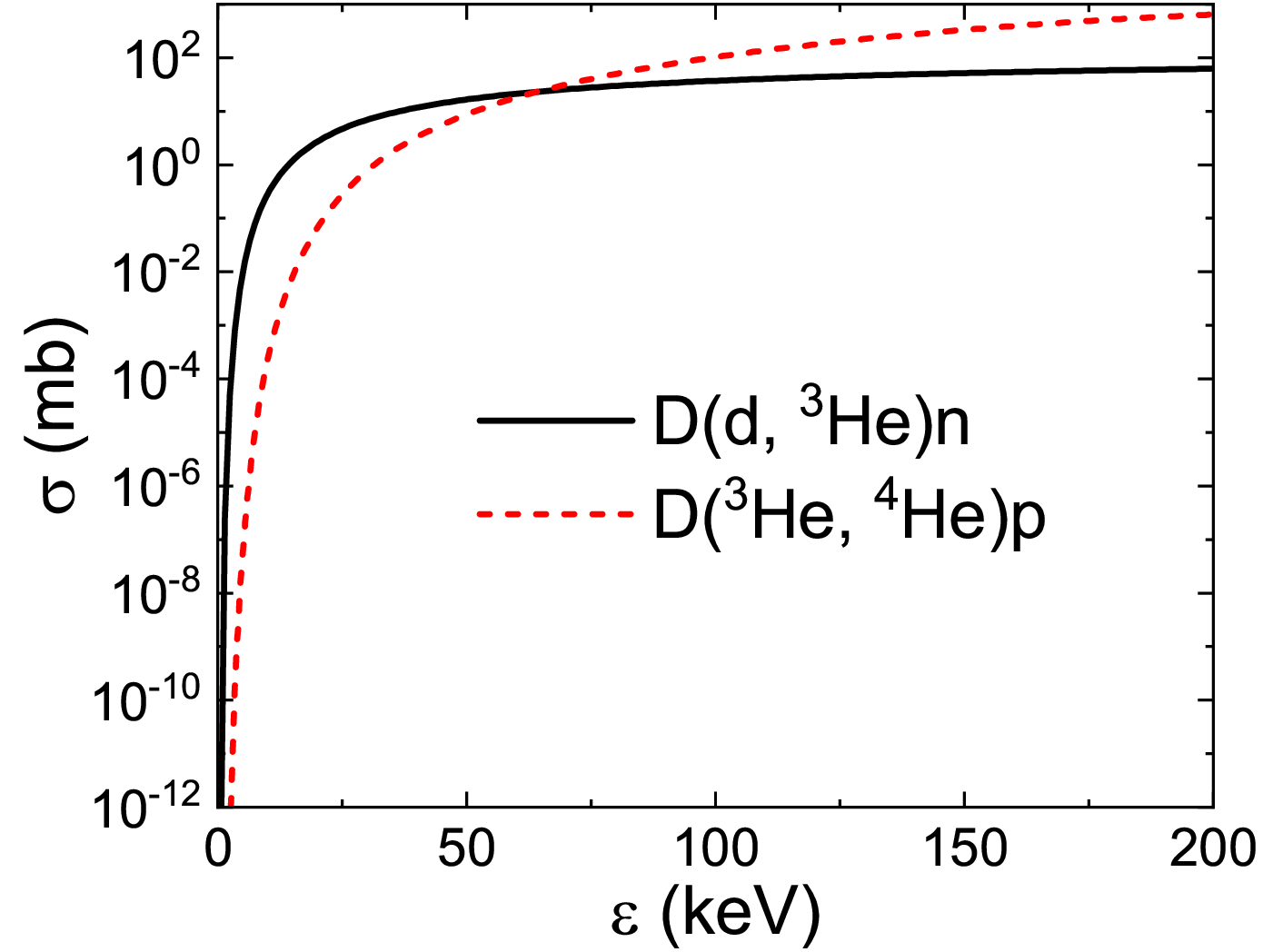}
\caption{(Color online) Cross sections of D+D $\rightarrow$ n+$^3$He and D+$^3$He $\rightarrow$ p+$^4$He channels as a function of the total kinetic energy $\epsilon$ in the C.M. frame of the collision from Ref.~\cite{Bos92}.}\label{fig1}
\end{figure}

In the simulation of Coulomb explosion for deuterium clusters, we set the deuteron number density $\rho_0=4.9 \times 10^{10} $ $\mu$m$^{-3}$ uniformly distributed inside clusters as in Ref.~\cite{Zha17}. The average deuteron density over the whole box system ranges from $10^7$ to $10^8$ $\mu$m$^{-3}$~\cite{Ban13} for different simulation events. To reach a similar condition as in Ref.~\cite{Ban13}, we assume cold deuterons and cold $^3$He uniformly distributed within the box with a density of $\rho^{}_D=2.26 \times 10^7$ $\mu$m$^{-3}$ and $\rho^{}_{^3He}=1.50 \times 10^7$ $\mu$m$^{-3}$, respectively, and they are treated only as background particles and do not affect the dynamics of hot deuterons from Coulomb explosion. We consider neutron production from collisions between hot deuterons and those between hot and cold deuterons in the D(d,n)$^3$He channel, and proton production from collisions between hot deuterons and cold $^3$He in the $^3$He(d,p)$^4$He channel. The D(d,p)T channel is neglected, since only energetic protons from $^3$He(d,p)$^4$He were measured experimentally. These inelastic channels are incorporated by using the stochastic method commonly used in simulations of heavy-ion collisions. For collisions between hot deuterons, the collision probability in a time interval $\Delta t$ and a local cell with the volume $(\Delta x)^3$ can be expressed as~\cite{Xu05}
\begin{equation}\label{Psto2}
P_{hot-hot} = v_{rel} \sigma^{}_{DD} \frac{\Delta t}{(\Delta x)^3},
\end{equation}
where $v_{rel}=|\vec{v}_1-\vec{v}_2|$ is the relative velocity with $\vec{v}_{1(2)}$ being the velocity of the first (second) deuteron, and $\sigma^{}_{DD}$ is the fusion cross section for the D(d,n)$^3$He channel. The above collision probability guarantees that the collision number per unit time per unit volume is exactly the reaction rate $\langle v_{rel} \sigma^{}_{DD} \rangle$, with $\langle ... \rangle$ being the average in local phase space. For collisions between a hot deuteron and background cold deuterons or $^3$He at rest, the collision probability in a time interval $\Delta t$ can be written as
\begin{equation}\label{Psto3}
P_{hot-cold} = v^{}_{D} \sigma^{}_{DD/D^3He} \Delta t \rho^{}_{D/^3He},
\end{equation}
where $v^{}_{D}$ is the velocity of the hot deuteron, and $\sigma^{}_{DD}$ and $\sigma^{}_{D^3He}$ are the fusion cross sections for the D(d,n)$^3$He and $^3$He(d,p)$^4$He channels, respectively. We use $\Delta t = 5$ fs and $\Delta x=10 \Delta r$ for evaluating the reaction probability. While the final states of particles that experience inelastic collisions can be treated in a standard way as in Ref.~\cite{Ber88}, they are irrelevant to the dynamics, since the numbers of produced particles are extremely small. We adopt the fusion cross sections of D(d,n)$^3$He and $^3$He(d,p)$^4$He in Ref.~\cite{Bos92}, and their energy dependencies are displayed in Fig.~\ref{fig1}. Due to the relatively small cross sections with respect to the plasma density, these fusion reactions are perturbative to the dynamics. Therefore, instead of producing neutrons or protons with the small probability $P$ according to Eqs.~(\ref{Psto2}) and (\ref{Psto3}), we generate neutrons and protons in each inelastic collision but these nucleons carry the weight of $P$. Due to the large numbers of inelastic collisions, the statistical error of nucleon yield calculated from $\sqrt{\sum P^2}$ is actually very small.%It is seen that the cross section of $^3$He(d,p)$^4$He is much smaller than that of D(d,n)$^3$He, and the total kinetic energy $\epsilon$ is expected to be higher for D+D pairs than for D+$^3$He pairs. It is expected that energetic deuterons affect significantly the weighted yield ratio as in Ref.~\cite{Ban13}.

\section{Results and discussions}

\begin{figure}[t]
\includegraphics[scale=0.3]{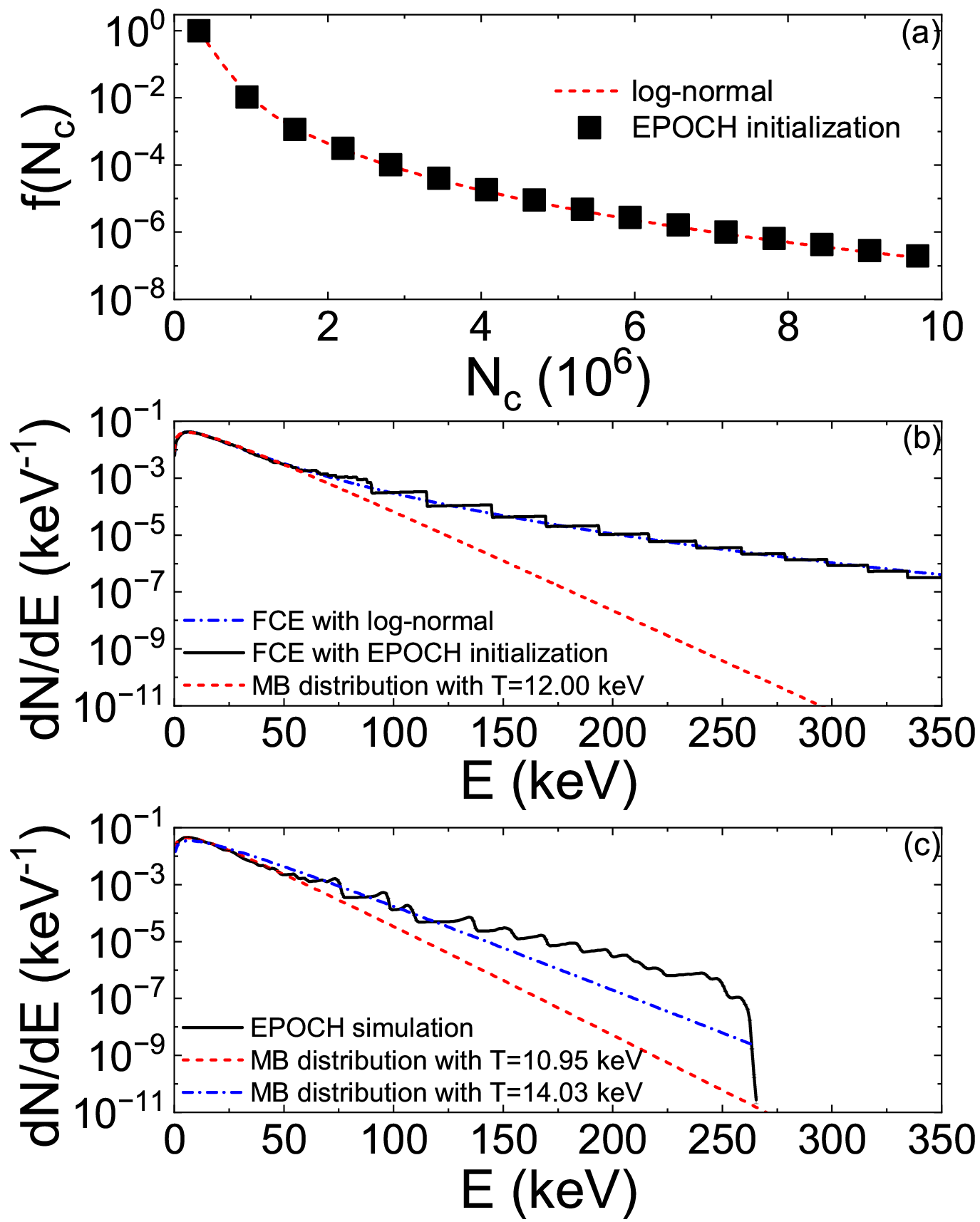}
\caption{(Color online) (a) Comparison of the exact log-normal distribution with the distribution of deuteron numbers in deuterium clusters in the initialization of EPOCH simulations; (b) Final deuteron energy spectra from free Coulomb explosion (FCE) with exact log-normal distribution of deuteron numbers, FCE with EPOCH initialization, and the fitted Maxwell-Boltzmann (MB) distribution with $T=12.00$ keV; (c) Final deuteron energy spectrum from EPOCH simulation and the Maxwell-Boltzmann (MB) distribution with $T=10.95$ and 14.03 keV. All distributions are normalized to 1. }\label{fig2}
\end{figure}

In the initialization of EPOCH simulations, the deuteron numbers $N_c$ in deuterium clusters are sampled according to the log-normal distribution~\cite{Lew93,Men03,Mad04}
\begin{equation} \label{log-normal}
f(N_c) = \frac{1}{N_c w \sqrt{2\pi}} \exp\left[-\frac{(\ln N_c -\mu)^2}{2w^2} \right],
\end{equation}
where the parameters are chosen to be $\mu=10.884$ and $w=1.066$, in order to reproduce the similar effective temperature as in Ref.~\cite{Ban13}. For each simulation event, there are 10 deuterium clusters, and the $N_c$ values of 1000 deuterium clusters in total are sampled ranging from $10^4$ to $10^7$, which covers the Gamow windows of the fusion reactions considered in the present study. Since the probability for the appearance of large $N_c$ is very small, we artificially choose 13 events for which the largest deuteron numbers $N^{max}_c$ in deuterium clusters is uniformly sampled within the range from $2\times10^6$ to $1\times10^7$. For the $N_c$ values of other deuterium clusters in these 13 events as well as those in the rest 87 events, they are sampled normally according to Eq.~(\ref{log-normal}). The weights of the 13 special events are set according to $f(N^{max}_c)$ with proper normalization, while those of the 87 normal events are set as 1. Figure~\ref{fig2}(a) compares the exact log-normal distribution with the sampled one for the initialization of EPOCH simulations, and it is seen that the two distributions are almost on top of each other. The box systems of 100 events in total can be considered as different local parts of the plasma, and the average deuteron densities are different in each event due to the log-normal distribution of $N_c$. Table~\ref{T1} has listed the number of computational cells, the number of computational particles, the average number density of hot deuterons, and the Debye length for the event containing the largest cluster with $N_c \approx 10^7$ and that containing the smallest cluster with $N_c \approx 10^4$. While the number of computational cells is larger than the number of computational particles, the grid size $\Delta r$ is much smaller than the Debye length. The simulation is numerically accurate, as shown in Figs. 3 and 4 of Ref.~\cite{Zhu22} where theoretical limits for the Coulomb explosion of deuterium clusters were reasonably reproduced.

\begin{table}[h!]
\centering
\caption{The number of computational cells (NCC), the number of computational particles (NCP), the average number density of hot deuterons ($n_D$), and the Debye length ($\lambda_D$) for the event containing the largest cluster (MAX) and that containing the smallest cluster (MIN) in EPOCH simulations.}
\label{T1}
\renewcommand\arraystretch{1.}
\setlength{\tabcolsep}{2mm}
% \resizebox{80mm}{!}
\begin{tabular}{|c|cccc|}
\hline
Event & NCC & NCP & $n_D$ ($\mu$m$^{-3}$) & $\lambda_D$ ($\mu$m) \\
\hline
MAX & $372^3$ & $1.02\times10^7$ & $2.90\times10^8$ & $4.37\times10^{-2}$ \\
MIN & $421^3$ & $5.01\times10^5$ & $9.81\times10^6$ & $2.37\times10^{-1}$ \\
\hline
\end{tabular}
\end{table}

If a single deuterium cluster experiences the Coulomb explosion individually without interacting with other deuterium clusters and background electrons, the resulting energy spectrum of final deuterons originating from the electrostatic energy can be expressed as~\cite{Zwe02}
\begin{equation}
\frac{dN_D}{dE} = 4\pi \epsilon_0 \frac{3}{2e^3} \sqrt{\frac{3\epsilon_0E}{\rho_0}},
\end{equation}
and the maximum energy that can be reached is~\cite{Zwe02}
\begin{equation}
E_{max} = \frac{e^2 \rho_0 R_0^2}{3\epsilon_0},
\end{equation}
where $R_0=(3N_c/4\pi\rho_0)^{1/3}$ is the radius of the deuterium cluster. If the deuteron numbers $N_c$ in all deuterium clusters follow the log-normal distribution $f(N_c)$, the total energy spectrum can then be calculated from
\begin{equation} \label{dnde}
\frac{dN_D}{dE} = C \int^\infty_{N_c(E)} f(N_c) \sqrt{E} dN_c,
\end{equation}
where $C$ is the normalization constant, and
\begin{equation}
N_c(E) = \left( \frac{3E\epsilon_0}{e^2\rho_0}\right)^{3/2} \frac{4\pi \rho_0}{3}
\end{equation}
is the lower-limit of the deuteron number for the Coulomb explosion of the deuterium cluster to reach the energy $E$. In Fig.~\ref{fig2} (b), the dash-dotted line represents the resulting deuteron kinetic energy spectrum from the exact log-normal distribution according to Eq.~(\ref{dnde}), and the solid line represents that by replacing $f(N_c)$ in Eq.~(\ref{dnde}) with the discrete distribution from EPOCH initialization as in Fig.~\ref{fig2} (a). The two spectra agree with each other rather well, and the low-energy part representing the majority of the spectrum can be fitted by a MB distribution with the temperature of about $T=12.00$ keV. The above calculation represents the ideal case of free Coulomb explosion. EPOCH simulations take the interactions with other deuterium clusters and background electrons into account, and the final deuteron spectrum is shown in Fig.~\ref{fig2} (c). Due to these interactions, the spectrum is steeper than that from free Coulomb explosion, and the majority of the resulting spectrum averaged over 100 events is fitted by a MB distribution with a lower temperature of $T=10.95^{+0.06}_{-0.07}$ keV. In both Fig.~\ref{fig2} (b) and Fig.~\ref{fig2} (c), there are considerable amount of deuterons at higher energies away from the fitted thermalized distributions. The intermediate- and high-energy parts of the spectrum, which cover the Gamow windows for the productions of neutrons and protons, are expected to significantly affect the weighted yield ratio used to extract the effective temperature to be discussed in the following.

\begin{figure}[t]
\includegraphics[scale=0.3]{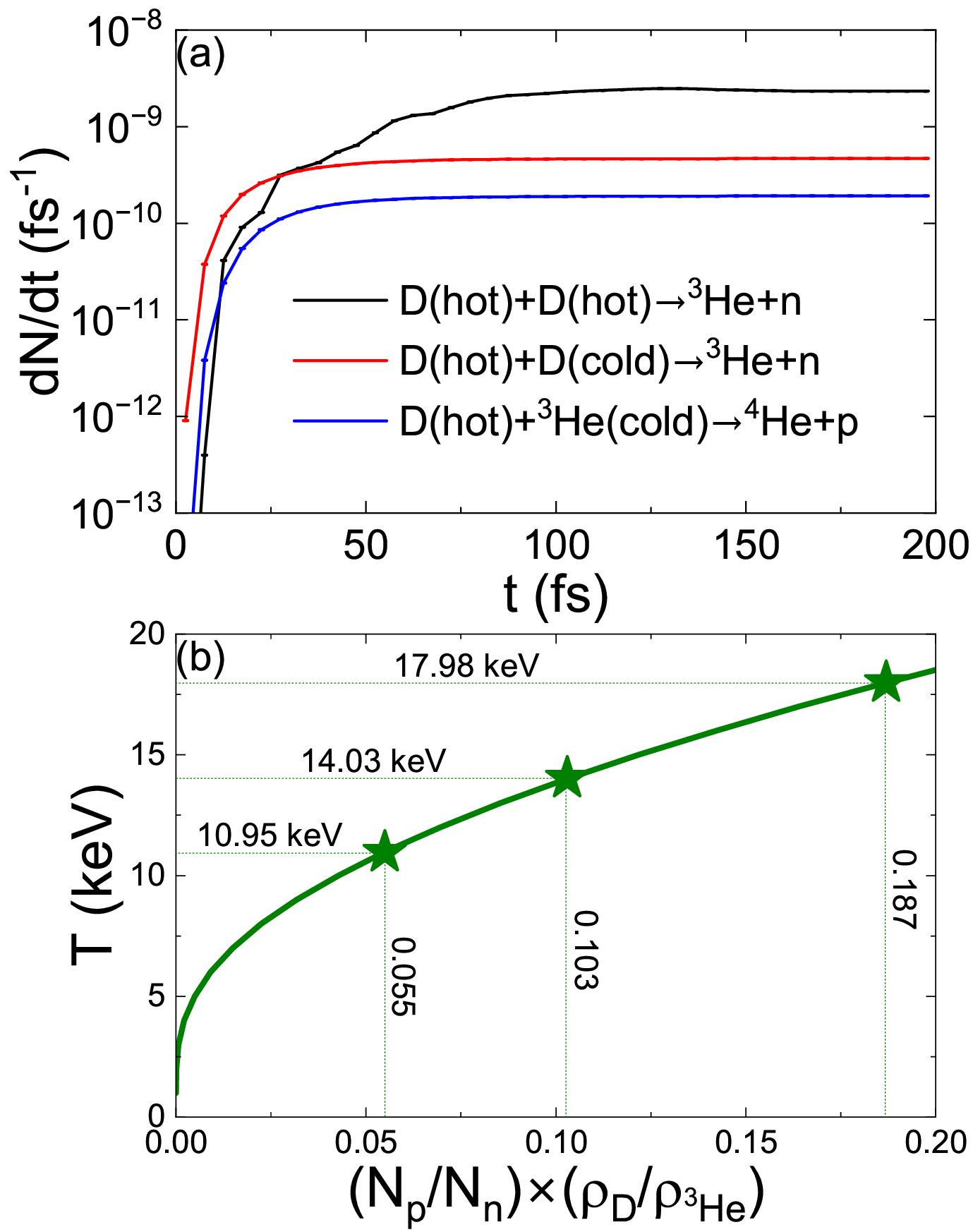}
\caption{(Color online) (a) Time evolutions of the three reaction rates per deuterium cluster for the production of neutrons and protons from EPOCH simulations; (b) Relation between the density weighted yield ratio and the temperature in the thermal model, with the star symbols representing different typical cases.}\label{fig3}
\end{figure}

The time evolutions of the fusion reaction rates of channels that produce neutrons and protons in the plasma from EPOCH simulations are shown in Fig.~\ref{fig3} (a). The reaction rates of hot-cold channels become equilibrated at about $t=50$ fs, due to the unchanged kinetic energy spectrum of deuterons afterwards. The reaction rate of hot deuteron-deuteron collisions reaches an equilibrium at a later time, after the plasma in each event becomes almost uniform. Times for equilibration of all channels are much short than the lifetime of the plasma in real experiments, which is of about a few hundred picoseconds. Figure~\ref{fig3}(b) shows the relation between the temperature and the density weighted proton-to-neutron yield ratio in the thermal model. Similar to Ref.~\cite{Ban13}, the relation can be expressed as
\begin{equation}\label{ratio}
\frac{N_p}{N_n} \frac{\rho^{}_D}{\rho^{}_{^3He}} = \frac{\langle \sigma^{}_{D^3He}v_{D} \rangle_{\frac{3}{5}kT}}{\frac{1}{2}\langle \sigma^{}_{DD}v_{rel} \rangle_{kT}+\langle \sigma^{}_{DD}v_{D} \rangle_{\frac{1}{2}kT}}.
\end{equation}
The numerator represents the reaction rate for collisions between hot deuterons and cold $^3$He, while the first and the second term in the denominator represent that for collisions between hot deuterons and between hot and cold deuterons, respectively, with the factor $1/2$ in the first term taking into account the degeneracy for identical particle collisions. As shown in the subscripts, different temperatures are used in evaluating the thermal average $\langle ... \rangle$ for different cases. The weighted yield ratio as in Eq.~(\ref{ratio}) increases with increasing temperature, as shown in Fig.~\ref{fig3} (b), and the resulting ratio of about $0.103$ from EPOCH simulations corresponds to a temperature of about $T=14.03$ keV. This temperature is considerably higher than that by fitting the deuteron kinetic energy spectrum, and it actually corresponds to a spectrum with more energetic deuterons as shown in Fig.~\ref{fig2} (c).

\begin{figure}[t]
\includegraphics[scale=0.3]{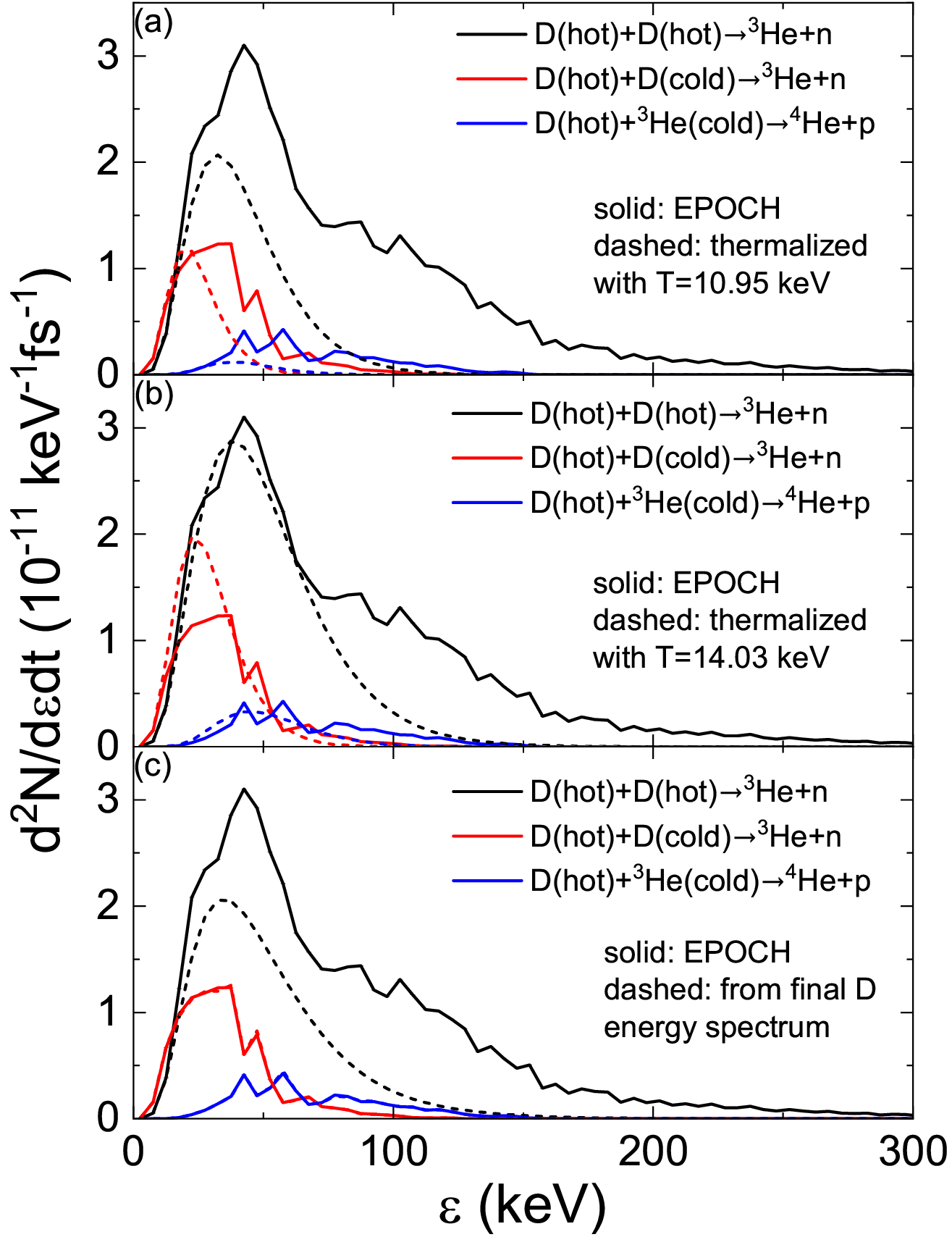}
\caption{(Color online) (a) C.M. energy distributions of the three reactions after the reaction rates become stable compared with the corresponding thermalized results with the temperature $T=10.95$ keV; (b) Same as panel (a) but compared with the corresponding thermalized results with the temperature $T=14.03$ keV; (c) Similar to panels (a) and (b) but compared with the results using the kinetic energy spectrum of deuterons from EPOCH simulations.}\label{fig4}
\end{figure}

We have also compared the C.M. energy distributions of the three reaction rates per deuterium cluster in the last 40 fs from EPOCH simulations with that from the thermal model at different temperatures in Fig.~\ref{fig4} (a) and Fig.~\ref{fig4} (b). In the thermal model, the results are obtained by assuming hot deuterons are uniformly distributed with their kinetic energies following a MB distribution at a certain temperature. It is seen that a temperature of about $T=10.95$ keV from fitting the kinetic energy spectrum of deuterons can only reproduce the distributions at very low energies but significantly underestimates the overall reaction rates, while the C.M. energy distributions from a temperature of about $T=14.03$ keV extracted from the weighted yield ratio agree much better with the results from EPOCH simulations. This is understandable, since the distributions around the Gamow windows of the fusion reactions as well as the nucleon yield attribute to intermediate- and high-energy deuterons rather than low-energy deuterons. The thermal model for $T=10.95$ keV corresponds to a density weighted yield ratio of about 0.055, as shown in Fig.~\ref{fig3} (b). Figure~\ref{fig4} (c) compares the C.M. energy distributions of the three reaction rates from EPOCH simulations with those from a modified thermal model, where the deuteron kinetic spectrum is taken exactly from the final result of EPOCH simulations as in Fig.~\ref{fig2} (c) instead of using a MB distribution. As expected, the C.M. energy distributions of hot-cold reactions are exactly the same as those from EPOCH simulations. However, the modified thermal model underestimates hot deuteron-deuteron collision numbers compared with EPOCH simulations. In all results from thermal model calculations in Fig.~\ref{fig4}, an averaged density from 100 EPOCH events is used, while in EPOCH simulations the densities are different in different events. For each EPOCH event, we find that the modified thermal model leads to almost the same hot deuteron-deuteron collision rate as EPOCH simulations, and events with higher (lower) densities lead to higher (lower) hot deuteron-deuteron reaction rates. The event-by-event density fluctuations lead to a higher hot deuteron-deuteron reaction rate from EPOCH simulations as shown in Fig.~\ref{fig4} (c), and this leads to nearly twice the neutron yield compared to that from the modified thermal model which uses the average density over 100 events. The modified thermal model gives a density weighted yield ratio of about 0.187 corresponding to a higher temperature of about 17.98 keV, as shown in Fig.~\ref{fig3} (b). We note that local density fluctuations intrinsically exist as long as the deuteron numbers in deuterium clusters follow a log-normal distribution. The above discussions show that the relation between the deuteron energy spectrum and the weighted yield ratio is non-trivial.

\section{Summary}

We have investigated the method of extracting the effective temperature from the weighted proton-to-neutron yield ratio through EPOCH simulations in box systems with periodic boundary conditions. From the initial log-normal distribution of deuteron numbers in deuterium clusters, the spectrum of final low-energy deuterons from Coulomb explosion can be fitted by a Maxwell-Boltzmann distribution, no matter whether Coulomb interactions between deuterium clusters are present or not, while there are more deuterons in the high-energy region compared to a thermal distribution. The fusion reactions, which are incorporated through the stochastic method, are dominated by intermediate- and high-energy deuterons that cover the Gamow windows. Therefore, the weighted yield ratio generally leads to a higher effective temperature compared to that from fitting the deuteron spectrum. The local density fluctuation enhances hot deuteron-deuteron collisions and produces more neutrons, and may further affect the weighted yield ratio and the extracted temperature. It is of interest to do simulations in a larger scale to see how large the effect of local density fluctuations is in the evolution of the whole plasma system. Our study helps to build a better understanding on the relation between properties of the plasma environment and the fusion reactions induced by high-intensity laser beam.

\begin{acknowledgments}
We acknowledge helpful discussions with Guo-Qiang Zhang and Aldo Bonasera. This work is supported by the Strategic Priority Research Program of the Chinese Academy of Sciences under Grant No. XDB34030000, the National Key Research and Development Program of China under Grants No. 2023YFA1606701, the National Natural Science Foundation of China under Grant Nos. 12235003 and 11922514, and the Fundamental Research Funds for the Central Universities.
\end{acknowledgments}

\end{document}